\documentstyle[a4,12pt,psfig]{article} 
\begin{document}
\thispagestyle{empty}
\begin{flushright}
  LU TP 02-32 \\
  September 2002
\end{flushright}
\vskip 2.5true  cm
\begin{center}
{\Large\bf 
Integral equation for spin dependent unintegrated parton distributions incorporating 
double $ln^2(1/x)$ effects at low $x$} 
\vskip .5true cm 
\vskip 1.5true cm
{\large\bf Jan Kwieci\'nski$^1$ and Martin Maul$^2$}
\\[1cm]
{$^1$The Henryk Niewodnicza\'nski 
  Institute of Nuclear Physics, 
Department of Theoretical Physics, 
Radzikowskiego 152, 31-342 Cracow, Poland 
\\
$^2$ Department of Theoretical Physics, Lund University,\\
S\"olvegatan 14A, S - 223 62 Lund, Sweden}\\
\vskip .5true cm
\end{center}
\vskip 1.5true cm
\begin{abstract}
\noindent
In this paper we derive  an integral equation for
the evolution of unintegrated (longitudinally) polarized quark and gluon 
parton distributions.  The conventional CCFM framework is modified at small 
$x$ in order to incorporate the QCD expectations concerning the 
 double $\ln^2(1/x)$ resummation at low $x$ for the integrated distributions.  
 Complete Altarelli-Parisi splitting functions 
are included, that makes the formalism compatible with 
the LO Altarelli-Parisi evolution at large and moderately small values of $x$.  
The obtained modified polarized CCFM  equation is shown to be partially diagonalized by 
the Fourier-Bessel transformation.  Results of the numerical solution for this 
modifed polarized CCFM equation for the non-singlet quark distributions are presented.        
\end{abstract}
\newpage
\section{Introduction}
The basic, universal quantities which describe the inclusive cross-sections 
of hard processes within the QCD improved parton model are the scale dependent 
parton distributions.  These distributions depend upon the longitudinal 
momentum fraction $x$ and the hard scale $Q^2$ and correspond to  the integrals 
over transverse momentum $k_\perp$ of the so called {\it unintegrated} distributions 
describing the (scale dependent) $x$ and $k_\perp$ parton distributions.  
The unintegrated distributions are needed in the description of less inclusive 
quantities which are sensitive to the transverse momenta of the partons \cite{Dokshitzer:1980}
- \cite{Lundsmx:2002}.\\

The unintegrated parton distributions are described in perturbative QCD by the 
Catani, Ciafaloni, Fiorani, Marchesini (CCFM) equation 
\cite{Ciafaloni:1987ur,Catani:1989yc,Catani:1989sg} 
which is based on color coherence which implies angular ordering. 
It embodies in a unified way the (LO) Altarelli-Parisi  evolution at large and moderately small 
values of $x$ with the BFKL dynamics at small $x$.\\

  The CCFM equation which was originally  formulated for unpolarized parton densities 
has recently been generalized to spin dependent unintegrated parton distributions 
\cite{Maul:2001uz}.  
Novel feature of the spin dependent case is its different structure 
at small $x$ and in particular the absence of the non-eikonal (or non-Sudakov) form-factors. 
We include here also the (spin dependent) quark distributions, leading first to a system of 
CCFM equations for unintegrated 
spin dependent quark and gluon distributions.    
We shall then extend the analysis of Ref. \cite{Maul:2001uz} along the 
following lines: 
\begin{enumerate}    
\item We shall incorporate theoretical QCD 
expectations concerning the double logarithmic small $x$ effects.  
We observe in particular that both the angular ordering constraint and the kernels 
of the corresponding system of the CCFM equations have to be appropriately modified. 
\item We shall include complete splitting functions $P_{ab}(z)$  and not only their 
singular and finite parts in the limit $z\rightarrow 1$ and $z\rightarrow 0$ respectively. 
\item We shall explore the relative simplicity of the system of the modified polarized CCFM equations  due to 
absence of non-eikonal form-factors and utilize the transverse coordinate representation 
of those equations.  The transverse coordinate $b_\perp$ is related to the transverse momentum 
of the partons through the Fourier-Bessel transformation and the system of the CCFM equation 
can be partially diagonalized by this transformation.
\end{enumerate}
The content of our paper is as follows: in the next section we recall the original 
formulation of the CCFM equation developed in \cite{Maul:2001uz} and include
the polarized quark distributions, while in 
sections 3 and 4 we formulate the modifications of the CCFM framework, which will 
incorporate the QCD expectations concerning the double $\ln^2(1/x)$ resummation 
at low $x$ and the complete Altarelli-Parisi evolution of integrated densities  in 
moderately small and large values of $x$.  We observe that the modifed CCFM equations, including their extension 
discussed in sections 4 and 5 can be partially diagonalized by the Fourier-Bessel 
transformation.   Section 5 is devoted to the numerical analysis 
of the modified CCFM equations and for simplicity we limit ourselves to the non-singlet quark distributions.  
Finally, in Section 6 we summarize  our main results and give our conclusions.       
    
\section{The original CCFM formulation for longitudinally polarized
unintegrated parton distributions}
\begin{figure}
\centerline{\psfig{figure=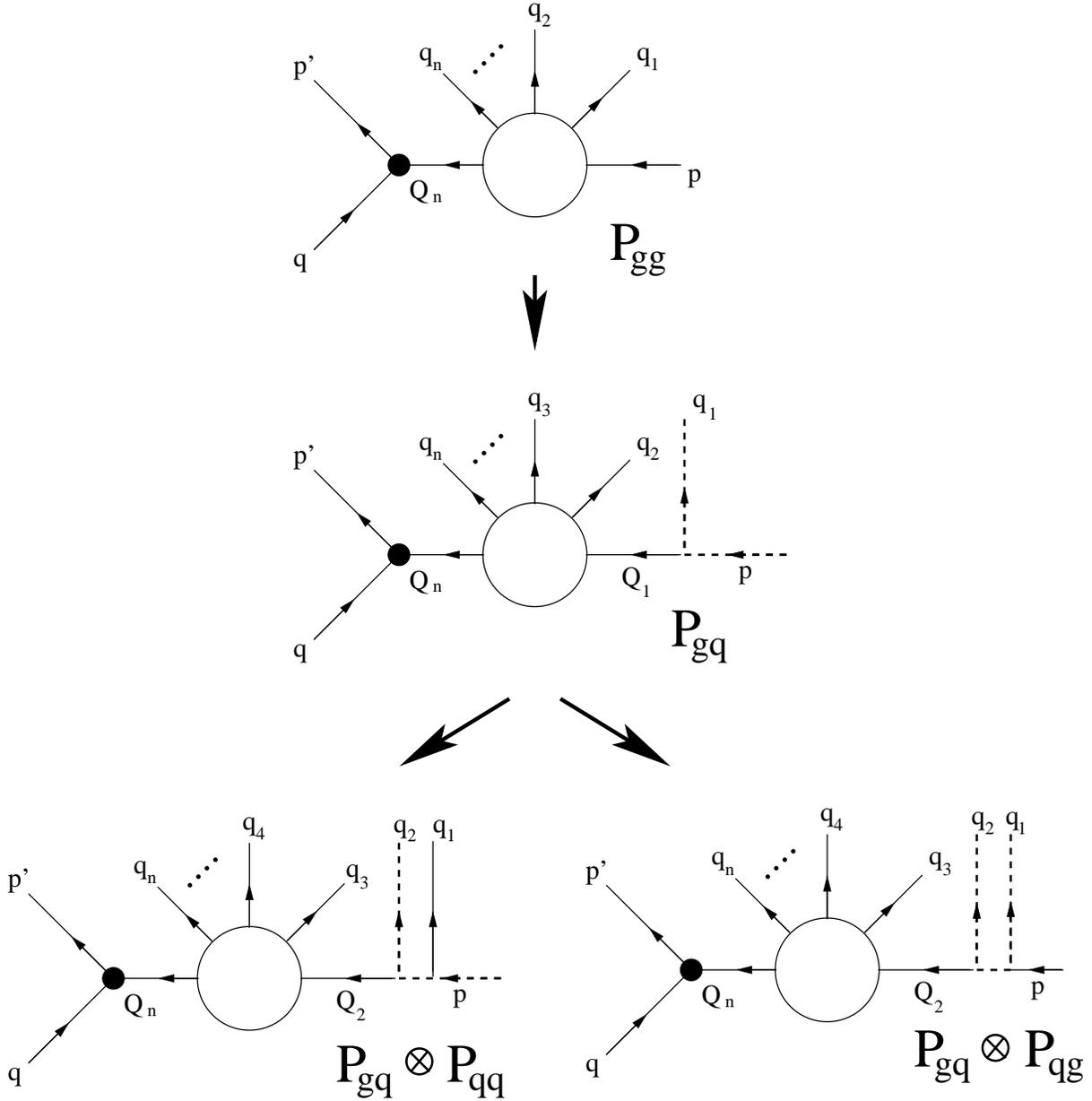,width=16cm}}
\caption{Construction of the polarized CCFM equation for quarks
from the one for gluons. The kernels for $qq$, $qg$ and $gq$ splitting
are subsequently constructed from the $gg$ soft emission by amending
the corresponding hard initial states. Here the solid line describes
gluons while the dashed line denotes quarks.}
\label{quarks}
\end{figure}

Along the lines of 
\cite{Ciafaloni:1987ur}\cite{Catani:1989yc}\cite{Catani:1989sg}
a version of the CCFM equation has been derived for the
longitudinally polarized gluon distribution \cite{Maul:2001uz}. 
For the corresponding expressions including quarks one has to 
take into consideration that the soft emission occurs predominantly
from a ladder of gluons because of their larger spin quantum number.
Therefore, quarks appear only as initial states playing automatically the
role of the hardest emission which enters in the Altarelli Parisi splitting
function, see Fig.~\ref{quarks}. Therefore, the arguments presented in 
\cite{Maul:2001uz} are valid in the case of quarks as well. The polarization
does not enter into the soft emission. Consequently, again a non-eikonal
form factor is absent and the eikonal form factor is identical to the one
in the unpolarized  case. Furthermore, in its original formulation the 
CCFM equation is an evolution equation valid only in the limits
$x\to0$ and $x\to 1$, and therefore, the Altarelli Parisi kernels
enter only in interpolated form, taking the limits $z\to0$ and $z\to 1$
into account. The complete polarized CCFM equations take the following 
general form:
\begin{eqnarray}
\Delta {\bf f} (x, k_\perp^2, Q^2) &=& 
\left(\Delta {\bf P^{\rm CCFM}} \otimes \Delta {\bf f}\right) 
(x, k_\perp^2, Q^2)
\nonumber \\
\Delta {f_k} (x, k_\perp^2, Q^2) &=& 
\left(\Delta { P^{\rm CCFM}_{qq}} \otimes \Delta { f_k}\right) 
(x, k_\perp^2, Q^2)\quad. 
\end{eqnarray}
Here we have used: 
\begin{equation}
\Delta {\bf f}= \left(
\begin{array}{c}
\Delta f_\Sigma \\ \\ \Delta f_g \end{array}\right), 
\quad {\rm and}\quad 
\Delta {\bf P^{\rm CCFM}}
= \left( \begin{array}{cc}
\Delta P^{\rm CCFM}_{qq}  & \Delta P^{\rm CCFM}_{qg} \\ \\ 
\Delta P^{\rm CCFM}_{gq}  & \Delta P^{\rm CCFM}_{gg}  
\end{array} \right)\;.
\end{equation}
Here $\Delta f_g$ is the unintegrated polarized gluon distribution function, $\Delta f_\Sigma$ the unintegrated 
polarized 
singlet quark parton distribution function and $ \Delta f_k$ denotes any polarized 
non-singlet combination of 
the quark parton distribution functions like the triplet $ (\Delta f_3)$ and the octet contribution ($\Delta f_8$).
For the convolution one has the structure:
\begin{eqnarray}
(\Delta P^{\rm CCFM}\otimes \Delta f)( x , k_\perp^2,Q^2)
&=&
\int_0^{2\pi} \frac{d \theta_{q_\perp}}{2\pi}
\int_{Q_0^2}^{\infty} \frac{d q_\perp^2}{ q_\perp^2}
\int_x^{1-Q_0/q} \frac{d z  }{ z } 
\Theta(Q- z  |{\bf q_\perp}|)
\nonumber \\ && \times
\Delta P^{\rm CCFM}(z,q_\perp^2, Q^2)
\Delta f( x / z ,{k_\perp'}^2, q_\perp^2) \;,
\end{eqnarray}
using $\vec k^{\prime}_{\perp} = \vec k_{\perp} + (1-z)\vec q_\perp$. In terms of the principles discussed
above one obtains for the CCFM splitting kernels:
\begin{eqnarray}
\Delta P_{gg}^{\rm CCFM}(z,q_\perp^2, Q^2)
&=&
\Delta^{(g)}_{\rm e}  ( Q^2,(z{\vec {q_\perp}})^2)      
\frac{\alpha_s(q_\perp^2(1-z)^2)}{2\pi}
\Delta P^{\rm AP0}_{gg }(z)
\nonumber \\
\Delta P_{gq}^{\rm CCFM}(z,q_\perp^2, Q^2)
&=&
\Delta^{(q)}_{\rm e}  ( Q^2,(z{\vec {q_\perp}})^2)      
\frac{\alpha_s(q_\perp^2(1-z)^2)}{2\pi}
\Delta P^{\rm AP0}_{gq }(z)
\nonumber \\
\Delta P_{qq}^{\rm CCFM}(z,q_\perp^2, Q^2)
&=&
\Delta^{(q)}_{\rm e}  ( Q^2,(z{\vec {q_\perp}})^2)      
\frac{\alpha_s(q_\perp^2(1-z)^2)}{2\pi}
\Delta P^{\rm AP0}_{qq }(z)
\nonumber \\
\Delta P_{qg}^{\rm CCFM}(z,q_\perp^2, Q^2)
&=&
\Delta^{(g)}_{\rm e}  ( Q^2,(z{\vec {q_\perp}})^2)      
\frac{\alpha_s(q_\perp^2(1-z)^2)}{2\pi}
\Delta P^{\rm AP0}_{qg }(z)\;.
\end{eqnarray}
The eikonal form factors have the form:
\begin{eqnarray}
\Delta_{\rm e}^{(g,q)}( q^2 , (zq)^2) &=&
\exp\left( - C_{A,F}\int_{(zq)^2}^{{ q}^2} \frac{d {q'}^2}{{q'}^2}
\int_0^{1-Q_0/{q'}} \frac{dz'}{1-z'} \frac{\alpha_s({q'}^2(1-z')^2)}{\pi} 
\right)\;.
\end{eqnarray}
and the interpolating Altarelli-Parisi kernels read then (c.~f.~\cite{Altarelli:1977zs}):
\begin{eqnarray}
\Delta P^{\rm AP0}_{gg }(z) &=& 2 C_A \frac{2-z}{1-z} 
\nonumber \\
\Delta P^{\rm AP0}_{qg }(z) &=& \left(z-\frac{1}{2}\right)
\nonumber \\
\Delta P^{\rm AP0}_{qq }(z) &=& C_F \frac{1+z^2}{1-z} 
\nonumber \\
\Delta P^{\rm AP0}_{gq }(z) &=& C_F(2-z)\;.
\end{eqnarray}
The fact that in the polarized case the CCFM splitting kernels are independent of 
$k_{\perp}$ allows
a factorization in terms of the Fourier-Bessel transformation:
\begin{eqnarray}
(\Delta P^{\rm CCFM}\otimes \Delta f)( x ,b_\perp^2,Q^2)
&=&
\int {d^2{\bf k_\perp}}e^{-i {\bf b_\perp \cdot k_\perp}}
(\Delta P^{\rm CCFM}\otimes \Delta f)( x ,k_\perp^2,Q^2)
\nonumber \\
&=&
\int_x^1 \frac{d z  }{ z } 
\int_{Q_0^2}^{Q^2/z^2} \frac{d q_\perp^2}{ q_\perp^2}
J_0((1-z){\bf| b_\perp|| q_\perp|})
\nonumber \\ && \times
\Delta P^{\rm CCFM}(z,q_\perp^2, Q^2)
\Delta \bar f( x / z ,b_\perp^2, q_\perp^2) \;.
\end{eqnarray}
One should note that $f(x,Q^2) = \bar f(x,0,Q^2)$ is just the $k_\perp$ integrated
parton distribution. 
\section{Modifications to make contact with ladder diagrams}
It may easily be observed that the angular ordering constraint which is embodied within the CCFM equation 
generates the double $\ln^2(1/x)$ terms  for the integrated parton distributions.  
%
The result concerning resummation of those terms turns out to be, however, different 
from the QCD expectations discussed in \cite{Bartels:1995iu}, 
\cite{Bartels:1996},\cite{Kirschner:1995},\cite{Kwiecinski:1999sk}. 
The double $ln^2(1/x)$ effects in QCD  are generated by the 
ladder and non-ladder bremsstrahlung diagrams.\\
In order to get the expected double logarithmic limit of the 
integrated distributions corresponding to the ladder diagrams contribution it 
is sufficient   
to replace the angular ordering constraint 
$\Theta(Q- z  |{\bf q_\perp}|)$ by 
the stronger constraint $\Theta(Q^2-zq_{\perp}^2)$ in the corresponding evolution 
equations for {\it integrated} distributions.  The latter just correspond to 
the Fourier transformed unintegrated evolution  equations 
at $b_\perp =0$.  We assume that this replacement can be done for arbitrary values of the transverse 
coordinate $b_\perp$. 
Due to this replacement we will obtain an equation which incorporates the known collinear (LO Altarelli-Parisi)
evolution for a not too small $x$ and the double logarithmic asymptotics for $x \ll 1$. 
However, the equation does not sum up the single $\ln (1/x)$ contributions as it was done by
the original CCFM/BFKL (unpolarized) equation. Having this point in mind we will henceforward
call the evolution equation for the polarized unintegrated parton distributions which we derive starting
from the CCFM formulation and where we are including our modfications a modified polarized CCFM
evolution equation.
Apart from this substitution in the argument of the theta function to include the double logarithmic contributions,
we shall also make the following 
modifications of the original CCFM equations proposed in \cite{Maul:2001uz}:   
%
\begin{enumerate}
\item The argument of $\alpha_s$ will  be set equal to $q_{\perp}^2$ instead 
of $q_{\perp}^2(1-z)^2$.  
\item The non-singular parts of the splitting function(s) 
will be included in the definition of the Sudakov form-factor(s). 
\item Following Ref. \cite{Kwiecinski:1999sk} we include complete splitting 
functions $P_{ab}(z)$ and not only their singular parts at $z=1$ and constant 
contributions at $z=0$.  
\item We represent the splitting functions $\Delta P_{ab}(z)$ as 
$\Delta P_{ab}(z)=\Delta P_{ab}(0)+ \Delta \bar P_{ab}(z)$ where 
$\Delta \bar P_{ab}(0)=0$.  Following \cite{Kwiecinski:1999sk} we shall 
multiply $\Delta P_{ab}(0)$ and $\Delta \bar P_{ab}(z)$ by 
$\Theta(Q^2-zq_{\perp}^2)$ and $\Theta(Q^2-q_{\perp}^2)$ respectively in the integrands 
of the corresponding integral equations.   Following  
the terminology of Ref. \cite{Kwiecinski:1999sk} we call the corresponding contributions 
to the evolution kernels  
the 'ladder' and 'Altarelli - Parisi ' contributions respectively.      
\item We shall 'unfold'  the eikonal  form factors in order to treat  
real emission and virtual correction terms on equal footing.  
\end{enumerate}
 
Using those prescriptions we get the following unfolded and modified polarized CCFM equations:

\begin{eqnarray}
\Delta f_g( x ,\vec k_\perp^2,Q^2)=\Delta \tilde f_g^0( x ,\vec k_\perp^2)
+\int{d^2 \vec q_\perp \over \pi q_\perp ^2}{\alpha_s(q_\perp ^2)\over 2 \pi}
\Theta(q_\perp ^2-Q_0^2)\int_0^1{dz\over z}
\nonumber
\end{eqnarray}

$$
\times\Bigg\{\Theta(z-x)
\Bigg[\left(12\Theta (Q^2-zq_\perp ^2)+6\Theta(Q^2-q_\perp ^2)\left({z\over 1-z}-2z\right)\right)
\Delta f_g( x/z ,\vec k^{\prime 2}_\perp,q_\perp ^2)
$$

$$
+\left({8\over 3}\Theta (Q^2-zq_\perp ^2)-{4\over 3}z\Theta(Q^2-q_\perp^2)\right) 
\Delta f_\Sigma( x/z ,\vec k^{\prime 2}_\perp,q_\perp ^2)\Bigg]
$$

\begin{equation}
-
z\Theta(Q^2-q_\perp ^2)\left({6\over 1-z}-{11\over 2}+{N_f\over 3}\right)\Delta 
f_g( x ,\vec k_\perp^2,q_\perp ^2)\Bigg\}
\label{ccfmmg}
\end{equation}

\begin{eqnarray}
\Delta f_{\Sigma}( x ,\vec k_\perp^2,Q^2)=\Delta \tilde f_{\Sigma}^0( x ,\vec k_\perp^2)
+\int{d^2 \vec q_\perp  \over \pi q^2}{\alpha_s(q_\perp ^2)\over 2 \pi}
\Theta(q_\perp ^2-Q_0^2)\int_0^1{dz\over z}
\nonumber 
\end{eqnarray}
$$
\times \Bigg\{\Theta(z-x)\Bigg[(-N_F\Theta (Q^2-zq_\perp ^2)+2zN_F\Theta(Q^2-q_\perp ^2))
\Delta f_g( x/z ,\vec k^{\prime 2}_\perp,q_\perp ^2)
$$

$$
+{4\over 3}\left(\Theta (Q^2-zq_\perp ^2)+{z+z^2\over 1-z}
\Theta(Q^2-q_\perp ^2)\right) 
\Delta f_{\Sigma}( x/z ,\vec k^{\prime 2}_\perp,q_\perp ^2)\Bigg]
$$

\begin{equation}
-z\Theta(Q^2-q_\perp ^2)
\left({8\over 3( 1-z)}-2\right)\Delta f_{\Sigma}( x ,\vec k_\perp^2,q_\perp ^2)\Bigg\}\;,
\label{ccfmmq}
\end{equation}
where we have put explicit numbers for the factors $C_A$ and $C_F$, and also introduced 
the singlet spin dependent unintegrated quark distributions: 
\begin{equation}
\Delta f_{\Sigma}= \Sigma_ {i=1}^{N_F} (\Delta f_{q_i} + \Delta f_{\bar q_i})\;.
\label{sigma}
\end{equation}
The 'non-singlet' quark distributions evolve as $\Delta f_{\Sigma}$ but  without the gluon 
contribution on the r.h.s. of the integral equation.\\

 In equations  (\ref{ccfmmg}) and 
(\ref{ccfmmq}) we set the upper limit of integration over $dz$ equal to 1 instead of $1-Q_0/q$ 
 since the integrands are free from singularities at $z=1$.  It should be noted that 
$\vec k^{\prime}_{\perp} = \vec k_{\perp} + (1-z)\vec q_\perp$. It should  also be noted that the 
inhomogeneous terms $\Delta \tilde f_g^0$ and $\Delta \tilde f_\Sigma^0$  in equations 
(\ref{ccfmmg}) and (\ref{ccfmmq}) do not contain the Sudakov form-factors. They  
can be chosen to have (for instance) the Gaussian form in $k_\perp$ normalized to unity and 
multiplied by the  input (integrated) distributions at the scale  $Q^2=Q_0^2$.  The latter 
could be taken from  one of 
the existing  (LO) QCD analysis of spin dependent parton distributions.  
To be precise the inhomogeneous terms are related to the starting distributions at the reference scale 
$Q_0^2$  in the 'single loop' approximation \cite{BRW,GMBRW} corresponding to the 
replacement $\Theta(Q^2-zq_{\perp}^2)$ by $\Theta(Q^2-q_{\perp}^2)$ in the integrals 
in equations (\ref{ccfmmg},\ref{ccfmmq}), since in general the integrals containing 
the function $\Theta(Q^2-zq_{\perp}^2)$ do not vanish at $Q^2=Q_0^2$.  
Parameterization of the driving term in terms of the parton distribution
may be regarded as a reasonable approximation, particularly in the 
region of large and moderately small values of $x$ which is dominated by the single 
loop approximation. \\   
   
Taking the Fourier-Bessel transformation on both sides of equations (\ref{ccfmmg}) and 
(\ref{ccfmmq}) we 
get the following equations for the distributions $ \bar f_g(x,b_\perp^2,Q^2)$ and 
$ \bar f_{\Sigma} (x,b_\perp^2,Q^2)$: 

\begin{eqnarray}
 \bar f_g( x ,b_\perp^2,Q^2)= \bar f_g^0( x ,b_\perp^2)
+\int{d q_{\perp}^2 \over \pi q_{\perp}^2}{\alpha_s(q_{\perp}^2)\over 2 \pi}
\Theta(q_{\perp}^2-Q_0^2)\int_0^1{dz\over z}
\nonumber 
\end{eqnarray}

$$
\times \Bigg\{J_0[b_\perp(1-z)q_{\perp}]\Theta(z-x)
\Bigg[\left(12\Theta (Q^2-zq_{\perp}^2)+6\Theta(Q^2-q_{\perp}^2)
\left({z\over 1-z}-2z\right)\right)
 \bar f_g( x/z ,b_\perp^2,q_{\perp}^2)
$$

$$
+\left({8\over 3}\Theta (Q^2-zq_{\perp}^2)-{4\over 3}z\Theta(Q^2-q_\perp^2)\right) 
\bar f_{\Sigma}( x/z ,b_\perp^2,q_{\perp}^2)\Bigg]
$$

\begin {equation}
-z\Theta(Q^2-q_\perp^2)
\left({6\over 1-z}-{11\over 2}+{N_f\over 3}\right) \bar f_g( x ,b_\perp^2,q_{\perp}^2)\Bigg\}
\label{ccfmbg}
\end{equation}
\begin{eqnarray}
 \bar f _{\Sigma}( x ,b_\perp^2,Q^2)=\bar f_{\Sigma}^0( x ,b_\perp^2)
+\int{dq_{\perp}^2 \over q_{\perp}^2}{\alpha_s(q_{\perp}^2)\over 2 \pi}
\Theta(q_{\perp}^2-Q_0^2)\int_0^1{dz\over z}
\nonumber 
\end{eqnarray}

$$
\times \Bigg\{J_0[b_\perp(1-z)q_{\perp}]\Theta(z-x)
\Bigg[(-N_F\Theta (Q^2-zq_{\perp}^2)+2zN_F\Theta(Q^2-q_{\perp}^2))
 \bar f_g( x/z ,b_\perp^2,q_{\perp}^2)
$$

$$
+{4\over 3}\left(\Theta (Q^2-zq_{\perp}^2)+{z+z^2\over 1-z}
\Theta(Q^2-q_{\perp}^2)\right) 
 \bar f_{\Sigma}( x/z ,b_\perp^2,q_{\perp}^2)\Bigg]
$$

\begin{equation}
-z\Theta(Q^2-q_\perp^2)
\left({8\over 3(1-z)}-2 \right)\bar f_{\Sigma}( x ,b_\perp^2,q_{\perp}^2)\Bigg\}\;,
\label{ccfmbq}
\end{equation}

where: 

\begin{eqnarray}
 \bar f_i(x,b_\perp^2,Q^2) &=& \int d^2{\vec k_\perp} \exp\left(-i\vec k_\perp \cdot b_\perp\right)
\Delta f_i(x,k_\perp^2,Q^2)
\nonumber \\ &=& 
2 \pi \int_0^\infty k_\perp dk_\perp J_0( k_\perp b_\perp) \Delta f_i(x,k_\perp^2,Q^2)\;,
\label{fb1}
\end{eqnarray}

and 

\begin{eqnarray}
\Delta  f_i(x,k_\perp^2,Q^2) &=& \int \frac{d^2 \vec b_\perp}{(2\pi)^2} 
 \exp\left(i\vec k_\perp \cdot b_\perp\right)
 \bar f_i(x,b_\perp^2,Q^2)
\nonumber \\
&=& \frac{1}{2\pi}  \int_0^\infty b_\perp db_\perp J_0( k_\perp b_\perp) \Delta f_i(x,b_\perp^2,Q^2)\;.
\label{fb2}
\end{eqnarray}
From a physical point of view one can interpret $b_\perp$ as an 'impact parameter' giving
the transverse distance of the partonic probe.
Then the following expressions are obtained:
\begin{eqnarray}
&&\overline{f}_g(x,b_\perp^2,Q^2) =
\overline{f}_{g}^0(x,b_\perp^2)
\nonumber \\
\nonumber \\
&&+ \int_x^1\frac{dz}{z}\int_{Q_0^2}^{Q^2/z} \frac{d q_\perp^2}{q_\perp^2}
\frac{\alpha_s(q_\perp^2)}{2\pi} \overline{f}_\Sigma(x/z,b_\perp^2,q_\perp^2)
 \frac{8}{3} J_0({\bf |b_\perp||q_\perp|}(1-z))
\nonumber \\
\nonumber \\
&&+ \int_x^1\frac{dz}{z}\int_{Q_0^2}^{Q^2/z} \frac{d q_\perp^2}{q_\perp^2}
\frac{\alpha_s(q_\perp^2)}{2\pi} \overline{f}_g(x/z,b_\perp^2,q_\perp^2)
12 J_0({\bf |b_\perp||q_\perp|}(1-z))
\nonumber \\
\nonumber \\
&& \qquad \qquad {\bf (ladder)}
\nonumber \\
\nonumber \\
\nonumber \\
&&+ \int_x^1\frac{dz}{z}\int_{Q_0^2}^{Q^2} \frac{d q_\perp^2}{q_\perp^2}
\frac{\alpha_s(q_\perp^2)}{2\pi} \overline{f}_\Sigma(x/z,b_\perp^2,q_\perp^2)
\left(- \frac{4}{3}z\right) J_0({\bf |b_\perp||q_\perp|}(1-z))
\nonumber \\
\nonumber \\
&&+\int_x^1\frac{dz}{z} \int_{Q_0^2}^{Q^2} \frac{d q_\perp^2}{q_\perp^2}
\frac{\alpha_s(q_\perp^2)}{2\pi}  6z\Bigg[\frac{ \overline{f}_g(x/z,b_\perp^2,q_\perp^2)
- \overline{f}_g(x,b_\perp^2,q_\perp^2)}{(1-z)} 
\nonumber \\ && \qquad   \qquad  \qquad \qquad  \qquad \qquad 
-2\overline{f}_g(x/z,b_\perp^2,q_\perp^2)\Bigg] 
J_0({\bf |b_\perp||q_\perp|}(1-z))
\nonumber \\
\nonumber \\
&&+ \int_{Q_0^2}^{Q^2} \frac{d q_\perp^2}{q_\perp^2}
\frac{\alpha_s(q_\perp^2)}{2\pi} \overline{f}_g(x,b_\perp^2,q_\perp^2)
\left[\frac{11}{2} - \frac{N_F}{3}+ 6 \ln(1-x) \right]
\nonumber \\
\nonumber \\
&& \qquad \qquad {\bf (Altarelli\;Parisi)}
\end{eqnarray}
\begin{eqnarray}
&&\overline{f}_\Sigma(x,b_\perp^2,Q^2) =
\overline{f}_{\Sigma}^0(x,b_\perp^2)
\nonumber \\
\nonumber \\
&&+ \int_x^1\frac{dz}{z}\int_{Q_0^2}^{Q^2/z} \frac{d q_\perp^2}{q_\perp^2}
\frac{\alpha_s(q_\perp^2)}{2\pi} \overline{f}_\Sigma(x/z,b_\perp^2,q_\perp^2)
\frac{4}{3} J_0({\bf |b_\perp||q_\perp|}(1-z))
\nonumber \\
\nonumber \\
&&- \int_x^1\frac{dz}{z}\int_{Q_0^2}^{Q^2/z} \frac{d q_\perp^2}{q_\perp^2}
\frac{\alpha_s(q_\perp^2)}{2\pi} \overline{f}_g(x/z,b_\perp^2,q_\perp^2)
 N_F J_0({\bf |b_\perp||q_\perp|}(1-z))
\nonumber \\
\nonumber \\
&& \qquad \qquad {\bf (ladder)}
\nonumber \\
\nonumber \\
\nonumber \\
&&+  \int_x^1\frac{dz}{z}\int_{Q_0^2}^{Q^2} \frac{d q_\perp^2}{q_\perp^2}
\frac{\alpha_s(q_\perp^2)}{2\pi}J_0({\bf |b_\perp||q_\perp|}(1-z))
\nonumber \\ && \qquad \qquad \times
\frac{4}{3}\left[\frac{(z+z^2) \overline{f}_\Sigma(x/z,b_\perp^2,q_\perp^2)
-2z \overline{f}_\Sigma(x,b_\perp^2,q_\perp^2)}{(1-z)}\right] 
\nonumber \\
\nonumber \\
&&+  \int_{Q_0^2}^{Q^2} \frac{d q_\perp^2}{q_\perp^2}
\frac{\alpha_s(q_\perp^2)}{2\pi} \overline{f}_\Sigma(x,b_\perp^2,q_\perp^2)
\left[2 + \frac{8}{3}\ln(1-x)\right]
\nonumber \\
\nonumber \\
&&+  \int_x^1\frac{dz}{z}\int_{Q_0^2}^{Q^2} \frac{d q_\perp^2}{q_\perp^2}
\frac{\alpha_s(q_\perp^2)}{2\pi} \overline{f}_g(x/z,b_\perp^2,q_\perp^2)
 2zN_F J_0({\bf |b_\perp||q_\perp|}(1-z))
\nonumber \\
\nonumber \\
&& \qquad \qquad {\bf (Altarelli\;Parisi)}
\end{eqnarray}
The contribution for the quark non-singlet part can be simply obtained from the expressions
for the singlet part leaving simply out all gluonic contributions:
\begin{eqnarray}
&&\overline{f}_{q\;NS}(x,b_\perp^2,Q^2) =
\overline{f}_{{q\;NS}}^0(x,b_\perp^2)
\nonumber \\
\nonumber \\
&&+ \int_x^1\frac{dz}{z}\int_{Q_0^2}^{Q^2/z} \frac{d q_\perp^2}{q_\perp^2}
\frac{\alpha_s(q_\perp^2)}{2\pi} \overline{f}_{q\;NS}(x/z,b_\perp^2,q_\perp^2)
\frac{4}{3} J_0({\bf |b_\perp||q_\perp|}(1-z))
\nonumber \\
\nonumber \\
&& \qquad \qquad {\bf (ladder)}
\nonumber \\
\nonumber \\
\nonumber \\
&&+ \int_x^1\frac{dz}{z}\int_{Q_0^2}^{Q^2} \frac{d q_\perp^2}{q_\perp^2}
\frac{\alpha_s(q_\perp^2)}{2\pi} 
 J_0({\bf |b_\perp||q_\perp|}(1-z))
\nonumber \\ && \qquad \qquad \times
\frac{4}{3}\left[ \frac{
(z+z^2)\overline{f}_{q\;NS}(x/z,b_\perp^2,q_\perp^2)
-2z \overline{f}_{q\;NS}(x,b_\perp^2,q_\perp^2)
}{(1-z)}\right] 
\nonumber \\
\nonumber \\
&&+ \int_{Q_0^2}^{Q^2} \frac{d q_\perp^2}{q_\perp^2}
\frac{\alpha_s(q_\perp^2)}{2\pi} \overline{f}_{q\;NS}(x,b_\perp^2,q_\perp^2)
\left[2+\frac{8}{3}\ln (1-x)\right]
\nonumber \\
\nonumber \\
&& \qquad \qquad {\bf (Altarelli\;Parisi)}
\nonumber \\
\end{eqnarray}

The expressions show that  except for the scale of $\alpha_s$ and the occurrence
of the Bessel function $J_0$ the expression match exactly the ones derived in 
\cite{Kwiecinski:1999sk} calculating ladder diagrams and combining this with the
Altarelli Parisi evolution. This shows the tight relationship that can be shown
between modified CCFM and the ladder contributions. 
\section{Inclusion of non-ladder diagrams}
There is a third contribution to the evolution of unintegrated parton distributions which is
not covered by the 'Altarelli - Parisi +ladder' approximation of the modified polarized 
CCFM equation,
these are the non-ladder bremsstrahlung
contributions. The method of implementing the non-ladder bremsstrahlung corrections in
general
into the double logarithmic resummation was proposed by Kirschner and Lipatov
\cite{Kirschner:1995}. For integrated polarized parton distributions
they have been implemented in Ref.~\cite{Kwiecinski:1999sk}.
The method developed  in Ref. \cite{Kirschner:1995} is based on the infrared 
equations for the partial waves $\bf F_{0,8}(\omega,\alpha_s)$:
\begin{equation}
{\bf F_{0,8}} = \left( \begin{array}{cc}
F_{0,8}^{qq} & F_{0,8}^{qg} \\ \\
F_{0,8}^{gq} &   F_{0,8}^{gg} 
\end{array} \right)\;.
\end{equation}
The infrared equations for the partial waves read:
\begin{equation}
{\bf F_0}(\omega,\alpha_s) =
 \frac{4\pi \alpha_s}{\omega}{\bf M_0}
-\frac{2\alpha_s }{\pi\omega^2}  {\bf F_8}(\omega,\alpha_s){\bf G_0} 
+\frac{1}{8\pi^2\omega} {\bf F_0^2}(\omega,\alpha_s)
\label{bff0}
\end{equation}
\begin{equation}
{\bf F_8}(\omega,\alpha_s)= 
 \frac{4\pi \alpha_s}{\omega}{\bf M_{8}}
+\frac{\alpha_s N}{2\pi\omega} \frac{d}{d\omega} {\bf F_8}(\omega,\alpha_s) 
+\frac{1}{8\pi^2\omega} {\bf F_8^2}(\omega,\alpha_s)\;,
\label{bff8}
\end{equation}
where the matrices ${\bf M_0}, {\bf M_8}$ and ${\bf G_0}$ are given by: 
\begin{equation}
{\bf M_8} = \left( \begin{array}{cc}
\frac{-1}{2N} & - \frac{N_F}{2} \\ \\
N              &   2N           \end{array} \right)
\end{equation}
\begin{equation}
{\bf M_0} = \left( \begin{array}{cc}
\Delta P_{qq}(0) & \Delta P_{qg}(0)   \\ \\
\Delta P_{gq}(0) &   \Delta P_{gg}(0)\end{array} \right)
\end{equation}
\begin{equation}
{\bf G_0} = \left( \begin{array}{cc}
 \frac{N^2-1}{2N} & 0 \\ \\
0              &   N           \end{array} \right)\;.
\end{equation}
The singlet partial wave matrix $\bf F_0$ is linked with the anomalous dimension 
matrix $\gamma_S^{RES}(\omega,\alpha_s)$ controlling the evolution of the moments 
of the integrated spin dependent distributions: 
\begin{equation}
{\bf F_0} = 8\pi^2 \gamma_S^{RES}(\omega,\alpha_s)\;.
\end{equation}
The anomalous dimension matrix corresponding to the solution of equation 
(\ref{bff0}) is given by: 
\begin{equation}
\gamma_S^{RES}(\omega,\alpha_s)= 
{\omega \over 2} \left(1-\sqrt{1-{2\alpha_s\over \pi \omega} \left( {{\bf M_0}\over 
\omega} -{{\bf F_8}(\omega,\alpha_s){\bf G_0} \over 2\pi^2 \omega^2}\right)}\right)\;.
\label{andimt}
\end{equation}
The anomalous dimension matrix contains the resummation of the double 
logarithmic $\ln^2(1/x)$ effects which  correspond to the sum of powers 
of $\alpha_s/\omega^2$ in the $\omega$ space. \\
It should be noted that the inhomogeneous term in the nonlinear equation 
(\ref{bff0}) which is proportional to ${\bf M_0}$ is also proportional to the 
kernel matrix defining the ladder diagram contributions in the double logarithmic 
approximation to 
the evolution equation for the (integrated)  parton distributions (cf. equations
 (\ref{ccfmmg}, \ref{ccfmmq}) at $b_\perp=0$).  
The fact that in equation (\ref{bff0}) 
 ${\bf M_0}$ appears in the 
inhomogeneous term while in equations (\ref{ccfmmg}, \ref{ccfmmq}) it 
appears in  the kernel matrix is linked with 
the fact that equation (\ref{bff0}) defines 
the $qq$-scattering amplitude in the $\omega$ representation
while 
equations  (\ref{ccfmmg}, \ref{ccfmmq}) define the parton distributions.  
The double logarithmic approximation of equations 
(\ref{ccfmmg}, \ref{ccfmmq}) at $b_\perp=0$  
would generate the anomalous dimension 
matrix $\gamma^{RES}_{\rm ladder}$ corresponding to ladder diagrams in the double logarithmic 
approximation which is given by:
 \begin{equation}
\gamma^{RES}_{\rm ladder}(\omega,\alpha_s)= 
{\omega \over 2} \left(1-\sqrt{1-{2\alpha_s {\bf M_0}\over\pi  
\omega^2}}\right)\;.
\label{andiml}
\end{equation}
It has been observed in  Ref. \cite{Kwiecinski:1999sk} that in order to get  
complete anomalous dimension given by equation (\ref{andimt}) one has to add the 
corresponding terms proportional to ${\bf F_8}(\omega,\alpha_s){\bf G_0}$ in the kernel matrix 
defining contribution of ladder diagrams. 
In the two-scale unintegrated
case one can simply add them by analogy
to the  Altarelli - Parisi  and ladder contribution
by inserting the factor  $J_0(b_\perp q_\perp (1-z))$.
The results are given below:
\begin{eqnarray}
&&\overline{f}_g(x,b_\perp^2,Q^2) =
\overline{f}_{g}^0(x,b_\perp^2)
\nonumber \\
\nonumber \\
&&+ \int_x^1\frac{dz}{z}\int_{Q_0^2}^{Q^2/z} \frac{d q_\perp^2}{q_\perp^2}
\frac{\alpha_s(q_\perp^2)}{2\pi} \overline{f}_\Sigma(x,b_\perp^2,q_\perp^2)
 \frac{8}{3} J_0({\bf |b_\perp||q_\perp|}(1-z))
\nonumber \\
\nonumber \\
&&+ \int_x^1\frac{dz}{z}\int_{Q_0^2}^{Q^2/z} \frac{d q_\perp^2}{q_\perp^2}
\frac{\alpha_s(q_\perp^2)}{2\pi} \overline{f}_g(x/z,b_\perp^2,q_\perp^2)
12 J_0({\bf |b_\perp||q_\perp|}(1-z))
\nonumber \\
\nonumber \\
&& \qquad \qquad {\bf (ladder)}
\nonumber \\
\nonumber \\
\nonumber \\
&&+ \int_x^1\frac{dz}{z}\int_{Q_0^2}^{Q^2} \frac{d q_\perp^2}{q_\perp^2}
\frac{\alpha_s(q_\perp^2)}{2\pi} \overline{f}_\Sigma(x/z,b_\perp^2,q_\perp^2)
\left(- \frac{4}{3}z\right) J_0({\bf |b_\perp||q_\perp|}(1-z))
\nonumber \\
\nonumber \\
&&+\int_x^1\frac{dz}{z} \int_{Q_0^2}^{Q^2} \frac{d q_\perp^2}{q_\perp^2}
\frac{\alpha_s(q_\perp^2)}{2\pi}  6z\Bigg[\frac{ \overline{f}_g(x/z,b_\perp^2,q_\perp^2)
- \overline{f}_g(x,b_\perp^2,q_\perp^2)}{(1-z)} 
\nonumber \\ && \qquad   \qquad  \qquad \qquad  \qquad \qquad 
-2\overline{f}_g(x/z,b_\perp^2,q_\perp^2)\Bigg] 
J_0({\bf |b_\perp||q_\perp|}(1-z))
\nonumber \\
\nonumber \\
&&+ \int_{Q_0^2}^{Q^2} \frac{d q_\perp^2}{q_\perp^2}
\frac{\alpha_s(q_\perp^2)}{2\pi} \overline{f}_g(x,b_\perp^2,q_\perp^2)
\left[\frac{11}{2} - \frac{N_F}{3}+ 6 \ln(1-x) \right]
\nonumber \\
\nonumber \\
&& \qquad \qquad {\bf (Altarelli\;Parisi)}
\nonumber \\
\nonumber \\
\nonumber \\
&&- \int_x^1\frac{dz}{z}\int_{Q_0^2}^{Q^2} \frac{d q_\perp^2}{q_\perp^2}
\frac{\alpha_s(q_\perp^2)}{2\pi} 
  J_0({\bf |b_\perp||q_\perp|}(1-z))
\left(\left[ \frac{\bf \tilde F_8}{\tilde \omega^2}\right](z) \frac{\bf G_0}{2\pi^2}\right)_{gq}
\overline{f}_\Sigma(x/z,b_\perp^2,q_\perp^2)
\nonumber \\
\nonumber \\
&&- \int_x^1\frac{dz}{z}\int_{Q^2}^{Q^2/z} \frac{d q_\perp^2}{q_\perp^2}
\frac{\alpha_s(q_\perp^2)}{2\pi} 
  J_0({\bf |b_\perp||q_\perp|}(1-z))
\left(\left[ \frac{\bf \tilde F_8}{\tilde 
\omega^2}\right]\left(\frac{q_\perp^2}{Q^2}z\right) 
\frac{\bf G_0}{2\pi^2}\right)_{gg}
\overline{f}_g(x/z,b_\perp^2,q_\perp^2)
\nonumber \\
\nonumber \\
&& \qquad \qquad {\bf (non-ladder)}
\nonumber \\
\end{eqnarray}
\begin{eqnarray}
&&\overline{f}_\Sigma(x,b_\perp^2,Q^2) =
\overline{f}_{\Sigma}^0(x,b_\perp^2)
\nonumber \\
\nonumber \\
&&+ \int_x^1\frac{dz}{z}\int_{Q_0^2}^{Q^2/z} \frac{d q_\perp^2}{q_\perp^2}
\frac{\alpha_s(q_\perp^2)}{2\pi} \overline{f}_\Sigma(x/z,b_\perp^2,q_\perp^2)
\frac{4}{3} J_0({\bf |b_\perp||q_\perp|}(1-z))
\nonumber \\
\nonumber \\
&&- \int_x^1\frac{dz}{z}\int_{Q_0^2}^{Q^2/z} \frac{d q_\perp^2}{q_\perp^2}
\frac{\alpha_s(q_\perp^2)}{2\pi} \overline{f}_g(x/z,b_\perp^2,q_\perp^2)
 N_F J_0({\bf |b_\perp||q_\perp|}(1-z))
\nonumber \\
\nonumber \\
&& \qquad \qquad {\bf (ladder)}
\nonumber \\
\nonumber \\
\nonumber \\
&&+  \int_x^1\frac{dz}{z}\int_{Q_0^2}^{Q^2} \frac{d q_\perp^2}{q_\perp^2}
\frac{\alpha_s(q_\perp^2)}{2\pi}J_0({\bf |b_\perp||q_\perp|}(1-z))
\nonumber \\ && \qquad \qquad \times
\frac{4}{3}\left[\frac{(z+z^2) \overline{f}_\Sigma(x/z,b_\perp^2,q_\perp^2)
-2z \overline{f}_\Sigma(x,b_\perp^2,q_\perp^2)}{(1-z)}\right] 
\nonumber \\
\nonumber \\
&&+  \int_{Q_0^2}^{Q^2} \frac{d q_\perp^2}{q_\perp^2}
\frac{\alpha_s(q_\perp^2)}{2\pi} \overline{f}_\Sigma(x,b_\perp^2,q_\perp^2)
\left[2 + \frac{8}{3}\ln(1-x)\right]
\nonumber \\
\nonumber \\
&&+  \int_x^1\frac{dz}{z}\int_{Q_0^2}^{Q^2} \frac{d q_\perp^2}{q_\perp^2}
\frac{\alpha_s(q_\perp^2)}{2\pi} \overline{f}_g(x/z,b_\perp^2,q_\perp^2)
 2zN_F J_0({\bf |b_\perp||q_\perp|}(1-z))
\nonumber \\
\nonumber \\
&& \qquad \qquad {\bf (Altarelli\;Parisi)}
\nonumber \\
\nonumber \\
\nonumber \\
&&-  \int_x^1\frac{dz}{z}\int_{Q_0^2}^{Q^2} \frac{d q_\perp^2}{q_\perp^2}
\frac{\alpha_s(q_\perp^2)}{2\pi} 
  J_0({\bf |b_\perp||q_\perp|}(1-z))
\left(\left[ \frac{\bf \tilde F_8}{\tilde \omega^2}\right](z) \frac{\bf G_0}{2\pi^2}\right)_{qq}
\overline{f}_\Sigma(x/z,b_\perp^2,q_\perp^2)
\nonumber \\
\nonumber \\
&&-  \int_x^1\frac{dz}{z}\int_{Q^2}^{Q^2/z} \frac{d q_\perp^2}{q_\perp^2}
\frac{\alpha_s(q_\perp^2)}{2\pi} 
  J_0({\bf |b_\perp||q_\perp|}(1-z))
\left(\left[ \frac{\bf \tilde F_8}{\tilde \omega^2}\right]\left(\frac{q_\perp^2}{Q^2}z\right)
 \frac{\bf G_0}{2\pi^2}\right)_{qg}
 \overline{f}_g(x/z,b_\perp^2,q_\perp^2)
\nonumber \\
\nonumber \\
&& \qquad \qquad {\bf (non-ladder)}
\nonumber \\
\end{eqnarray}
The contribution for the quark non-singlet part can be simply obtained from the expressions
for the singlet part leaving simply out all gluonic contributions:
\begin{eqnarray}
&&\overline{f}_{q\;NS}(x,b_\perp^2,Q^2) =
\overline{f}_{q\;NS}^0(x,b_\perp^2)
\nonumber \\
\nonumber \\
&&+ \int_x^1\frac{dz}{z}\int_{Q_0^2}^{Q^2/z} \frac{d q_\perp^2}{q_\perp^2}
\frac{\alpha_s(q_\perp^2)}{2\pi} \overline{f}_{q\;NS}(x/z,b_\perp^2,q_\perp^2)
\frac{4}{3} J_0({\bf |b_\perp||q_\perp|}(1-z))
\nonumber \\
\nonumber \\
&& \qquad \qquad {\bf (ladder)}
\nonumber \\
\nonumber \\
\nonumber \\
&&+ \int_x^1\frac{dz}{z}\int_{Q_0^2}^{Q^2} \frac{d q_\perp^2}{q_\perp^2}
\frac{\alpha_s(q_\perp^2)}{2\pi}  J_0({\bf |b_\perp||q_\perp|}(1-z))
\nonumber \\ && \qquad \qquad \times 
 \frac{4}{3}\left[\frac{
(z+z^2)\overline{f}_{q\;NS}(x/z,b_\perp^2,q_\perp^2)
-2z \overline{f}_{q\;NS}(x,b_\perp^2,q_\perp^2)
}{(1-z)}\right] 
\nonumber \\
\nonumber \\
&&+\int_{Q_0^2}^{Q^2} \frac{d q_\perp^2}{q_\perp^2}
\frac{\alpha_s(q_\perp^2)}{2\pi} \overline{f}_{q\;NS}(x,b_\perp^2,q_\perp^2)
\left[2+\frac{8}{3}\ln (1-x)\right]
\nonumber \\
\nonumber \\
&& \qquad \qquad {\bf (Altarelli\;Parisi)}
\nonumber \\
\nonumber \\
\nonumber \\
&&-  \int_x^1\frac{dz}{z}\int_{Q_0^2}^{Q^2} \frac{d q_\perp^2}{q_\perp^2}
\frac{\alpha_s(q_\perp^2)}{2\pi} 
  J_0({\bf |b_\perp||q_\perp|}(1-z))
\left(\left[ \frac{\bf \tilde F_8}{\tilde \omega^2}\right](z) \frac{\bf G_0}{2\pi^2}\right)_{qq}
\overline{f}_{q\;NS}(x/z,b_\perp^2,q_\perp^2)
\nonumber \\
\nonumber \\
&&-  \int_x^1\frac{dz}{z}\int_{Q^2}^{Q^2/z} \frac{d q_\perp^2}{q_\perp^2}
\frac{\alpha_s(q_\perp^2)}{2\pi} 
  J_0({\bf |b_\perp||q_\perp|}(1-z))
\left(\left[ \frac{\bf \tilde F_8}{\tilde \omega^2}\right]
\left(\frac{q_\perp^2}{Q^2}z\right) \frac{\bf G_0}{2\pi^2}\right)_{qq}
\overline{f}_{q\;NS}(x/z,b_\perp^2,q_\perp^2)
\nonumber \\
\nonumber \\
&& \qquad \qquad {\bf (non-ladder)}
\nonumber \\
\end{eqnarray}
Here $\Bigg(\left[ \frac{\bf \tilde F_8}{ \omega^2}\right](z)\Bigg)$ is the inverse Mellin
transformation of ${\bf F_8}(\omega)/\omega^2$. 
%
%
%
%
%
%
%
%
Derived for large $N$ and fixed $\alpha_s$, one can use an approximate
form \cite{Bartels:1995iu}:
\begin{equation}
\left[\frac{{\bf \tilde F_8}^{\rm Born}}{\omega}\right](z) = 2\pi \alpha_s {\bf M_8}\ln^2(z)\;,
\end{equation}
so that one gets in our case e.g.:
\begin{equation}
\left(\left[ \frac{\bf \tilde F_8}{\tilde \omega^2}\right](z) \frac{\bf G_0}{2\pi^2}\right)_{qq}
\approx -\frac{N^2-1}{4 \pi N^2} \alpha_s(q_\perp^2) \ln^2(z)\;.
\end{equation}
\section{Numerical studies}
\subsection{The input distributions}
For the polarized input distributions we are going to use the LO Standard Scenario parameterization given in
\cite{Gluck:2000dy}. The structure of the integrated input distributions
is a factor multiplied with the unpolarized input distributions given in \cite{Gluck:1998xa}.
As we have 'de-exponentiated' the eikonal form factor in the evolution equations, the
input distributions have to be $Q^2$ independent. This is different from the usual unpolarized
CCFM-input distributions, where one uses for the $Q^2$ dependence  the corresponding
eikonal form factor \cite{Jung:2000hk}. 
For the $k_\perp$ dependence of the 
input distributions a Gaussian Ansatz is common \cite{Jung:2000hk}. In this way the 
input distribution as defined at an input scale $Q_0^2$ have the following general scheme:
%
%
%
\begin{eqnarray}
\Delta \tilde f^0_i(x,k_\perp^2,Q^2) &=& \Delta p_i(x,Q_0^2)\exp\left(- \frac{k_\perp^2}{\sigma^2}\right) \frac{1}{\pi\sigma^2}
\nonumber \\
\Rightarrow \Delta \bar  f^0_i(x,b_\perp^2,Q^2) &=&
 \Delta p_i(x,Q_0^2) 
\exp\left( - \frac{b_\perp^2 \sigma^2}{4} \right)\;.
\end{eqnarray}
where $\Delta p_i(x,Q_0^2)$ are the input integrated spin dependent parton 
distributions at the reference scale $Q_0^2$.  
\subsection{Features of the evolution}
Setting $b_\perp = 0$ in Eqs.~(\ref{ccfmbq}) and (\ref{ccfmbg}), one obtains
evolution equations for  the integrated polarized parton distributions
equivalent to those  given in 
\cite{Kwiecinski:1999sk}.  Therefore, the $x$ and $Q^2$ dependence of the unintegrated parton 
distributions will be exactly the same as given by the Altarelli Parisi evolution equations 
supplemented by ladder and non-ladder contributions. 
In this way 
 the modified polarized CCFM equation as
discussed here is consistent with the standard evolution of the integrated polarized parton 
distributions. In fact the whole transverse momentum dependence is governed by the 
inclusion of the factor $J_0(b_\perp q_\perp (1-z))$. In principle, 
the only genuine new feature
of the modified CCFM equation  is the $k_\perp$ (or $b_{\perp}$) dependence.
At large and moderately small values of $x$ one can make the 'single-loop' approximation
\cite{BRW,GMBRW} corresponding to the replacement $\Theta(Q^2-zq_{\perp}^2)$
by just $\Theta(Q^2-q_{\perp}^2)$ in the 'ladder' contribution and to neglecting the
'non-ladder' contribution.  At $b_\perp =0$ the modified polarized CCFM equation in the 
single loop approximation equation reduces then
to the LO Altarelli-Parisi equation in the integral form. \\

\begin{figure}
\centerline{\psfig{figure=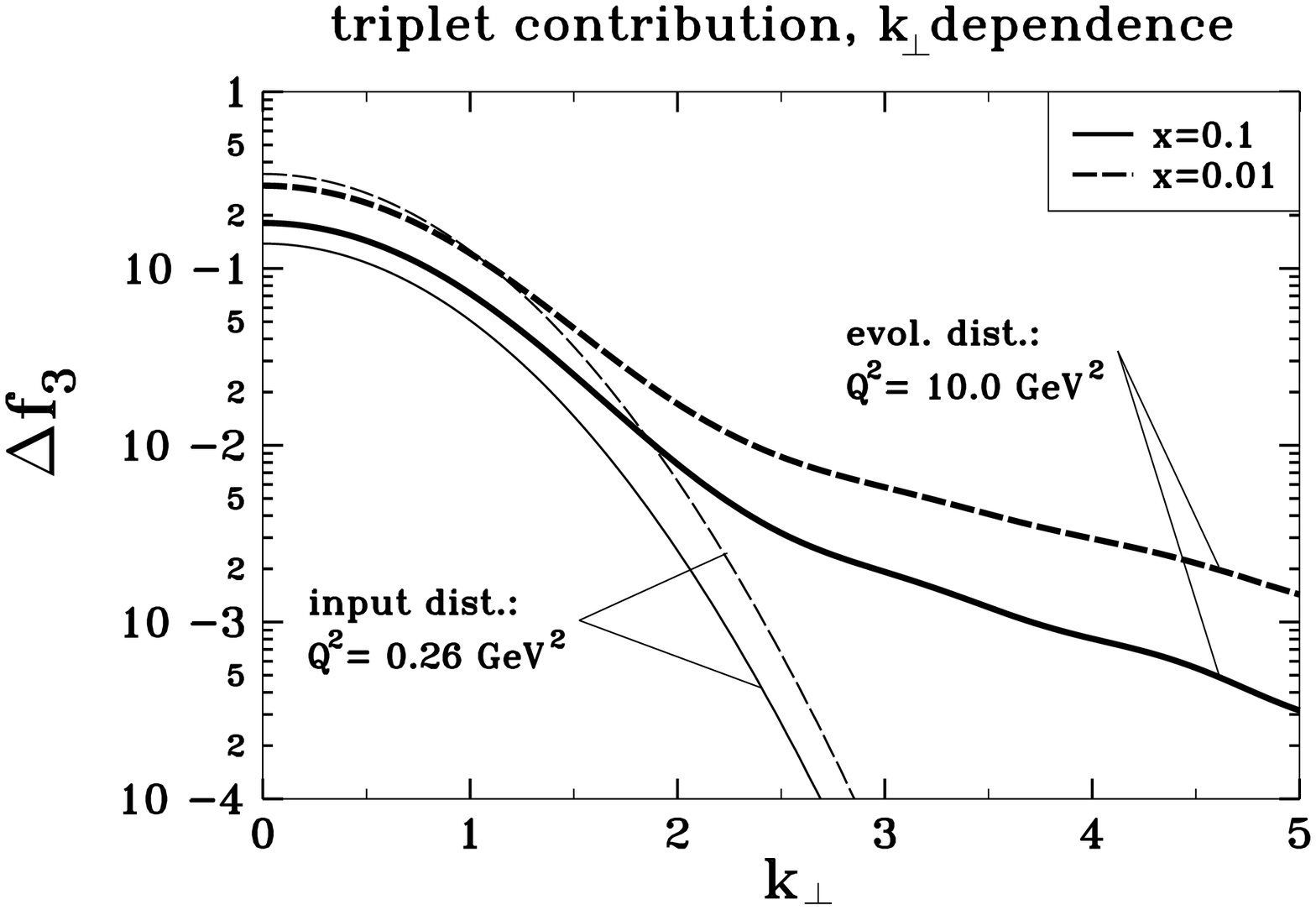,width=14cm}}
\centerline{\psfig{figure=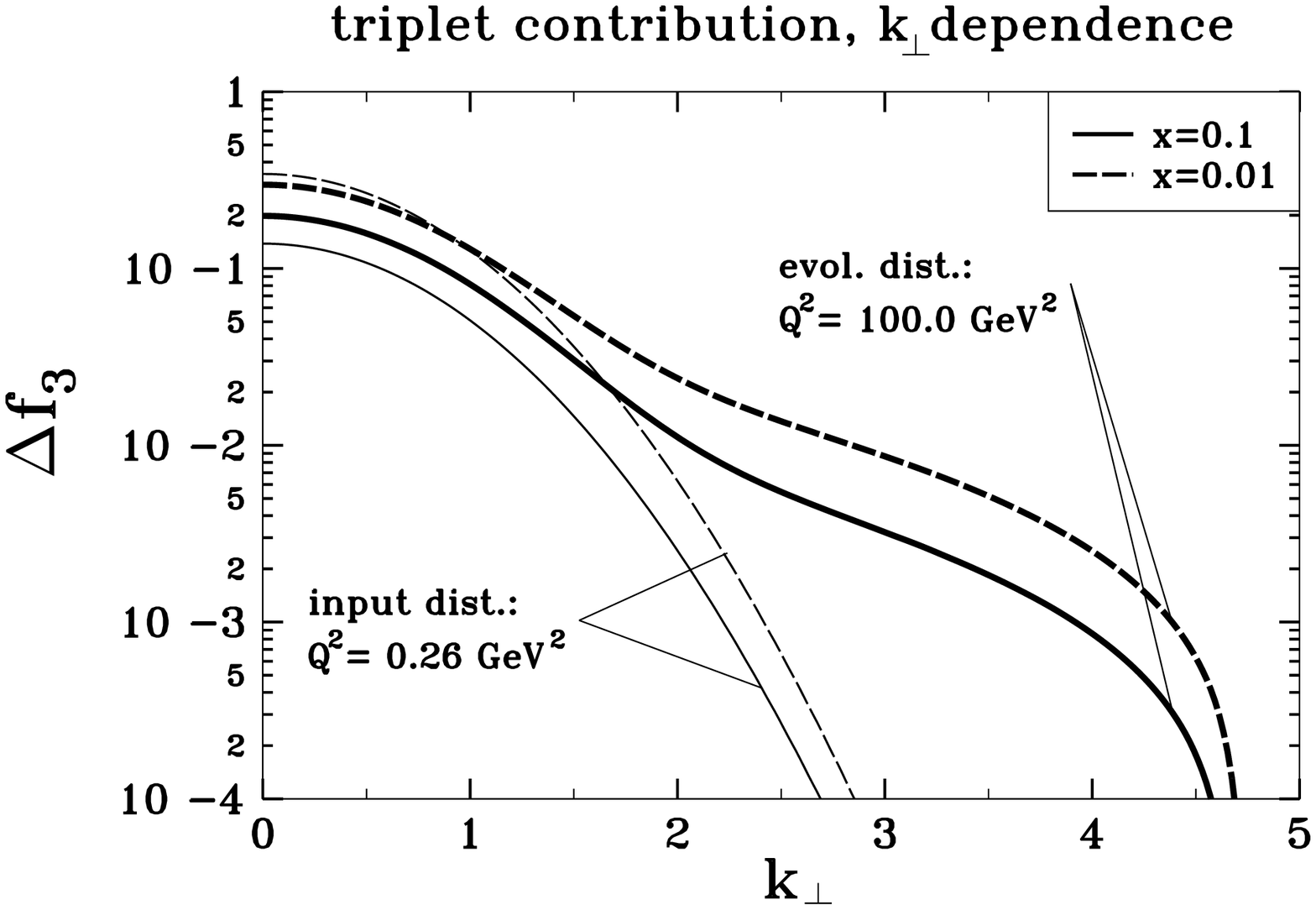,width=14cm}}   
\caption{Transverse momentum dependence for the triplet contribution 
$\Delta f_3$ = $\frac{1}{6} \left(\Delta u + \Delta \bar u-\Delta d - \Delta \bar d \right)$
of the full modified pCCFM evolution (approximate form: Altarelli Parisi + ladder + non-ladder). 
The thin lines show the input distributions GRSV LO Standard Scenario \cite{Gluck:2000dy}
for $Q^2_0 = 0.26\; {\rm GeV}^2$ for $x=0.1$ (solid) and $x=0.01$ (dashed), while the bold
lines show the same distribution evolved
to $Q^2= 10\;{\rm GeV}^2$ (top) and $Q^2= 100\;{\rm GeV}^2$ (bottom). It can be seen from 
the logarithmical scale that the evolution leads to a $k_\perp$ broadening away from the
Gaussian shape to a mere simple exponential decay.}
\label{ccfmk}
\end{figure}

Numerically, we perform
the evolution using the Chebyshev approximation technique as discussed in    
the Appendix.
The calculation is very time consuming, therefore, we take only 8 polynomials in all cases into
account. As the Bessel function $J_0$ is oscillating the corresponding integration is
numerically quite problematic. As a pragmatical solution the integration is only performed up
to the fourth zero. We have checked that this procedure provides stable results. As the mathematical
structure is the same for singlet and non-singlet contributions we can restrict ourselves for 
the discussion of the $k_\perp$ dependence to the simple non-singlet case. In Fig.~\ref{ccfmk}
we show the evolution of the $k_\perp$ dependence for  the triplet contribution 
$\Delta f_3$ = $\frac{1}{6} \left(\Delta u + \Delta \bar u-\Delta d - \Delta \bar d \right)$.
The input distributions at $Q^2_0 = 0.26 \;{\rm GeV}^2$ taken from the GRVS LO Standard Scenario set
\cite{Gluck:2000dy} 
are compared to the evolved distributions
at $Q^2 = 10.0 \;{\rm GeV}^2$ and $Q^2 = 100.0 \;{\rm GeV}^2$. The width of the initial transverse momentum dependence 
$\sigma$ has been chosen to be 1 GeV. For the simulation the full content of the equation,
the Altarelli-Parisi, ladder and non-ladder contributions all have been included.
It is seen that due to the evolution the $k_\perp$ dependence
is broadening away from a Gaussian behavior to a more purely exponential decay. Such a feature
has already been seen in the purely gluonic formulation of the genuine polarized CCFM equation
as discussed in \cite{Maul:2001uz}.
\begin{figure}
\centerline{\psfig{figure=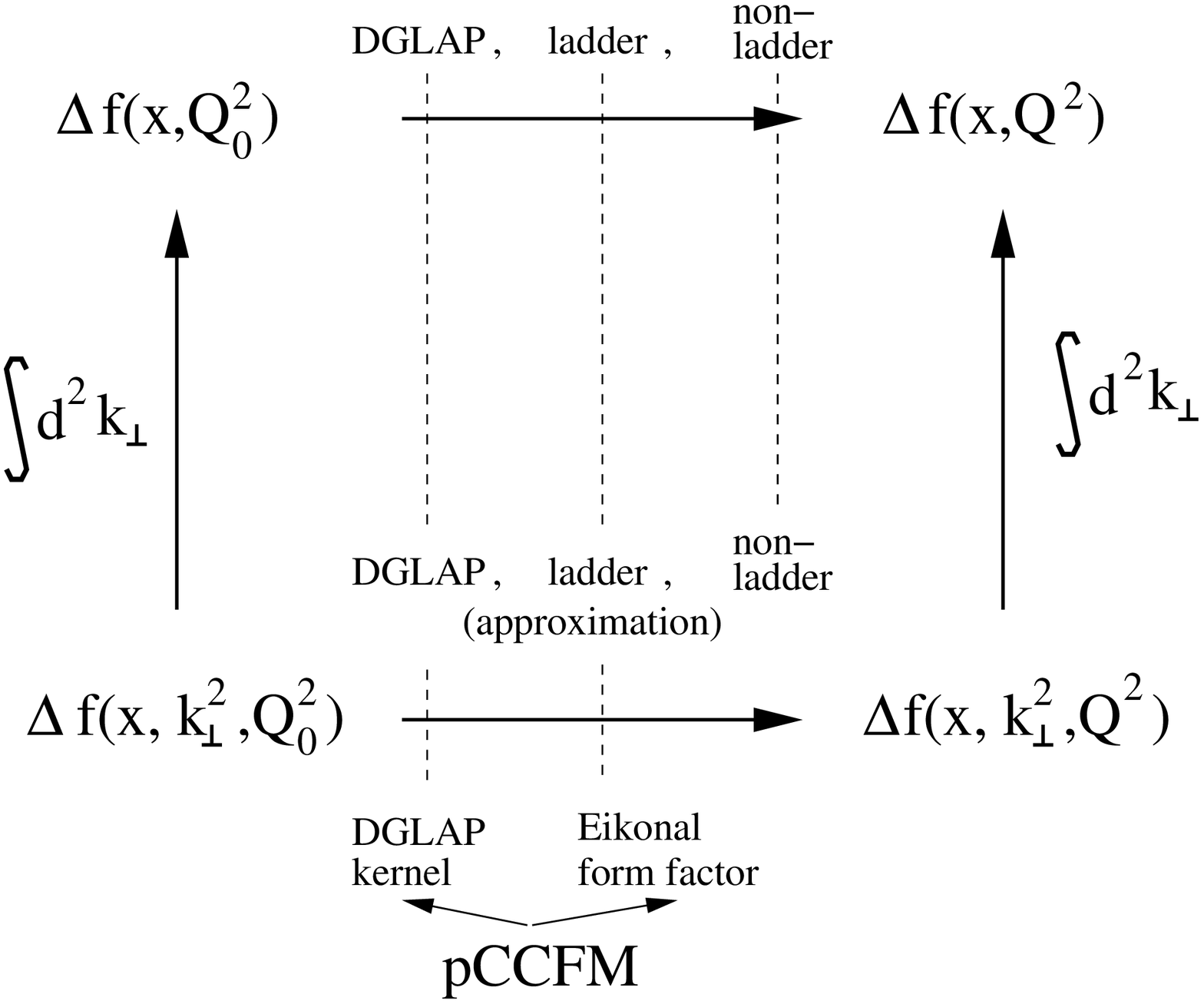,width=12cm}}
\caption{Relationship between the pCCFM evolution and its modified approximate form
for unintegrated parton distributions on the one hand 
and the evolution of integrated parton distributions on the other hand.}
\label{relation}
\end{figure}

\section{Summary and conclusions}
 In this paper we have derived an integral equation of the unintegrated polarized parton distributions
starting from a genuine polarized CCFM (pCCFM) formulation and including some modifications incorporating 
the known collinear  (LO Altarelli-Parisi) evolution  for a not too small $x$ and the double log asymptotics
for $x \ll 1$. This evolution equation which we call  modified polarized CCFM  equation does not sum up
the single $\ln(1/x)$ as it was done by the CCFM/BFKL unpolarized equation. An inclusion of the 
single log contributions would be very interesting, but is unfortunately beyond the scope of this article.
The  modified CCFM equation yields an approximate description, which contains contributions
of the type of 'Altarelli Parisi', 'ladder' and 'non-ladder' and  which correspond to respective
contributions in the integrated case.
We have utilized the fact that these equations can be diagonalized using the Fourier-Bessel transformation
with the 'impact parameter' $b_\perp$. The nice feature of this transformation is that by simply
setting  $b_\perp=0$ one obtains already the corresponding expressions for the integrated parton
distributions. In fact, the difference between both is only a factor $J_0(b_\perp q_\perp (1-z))$.
As to the $b_\perp$ dependence the mathematical structure is the same for the non-singlet and
singlet contributions to the unintegrated parton distributions. Using the technique of the
Chebyshev approximation we have performed the evolution for the triplet distribution $\Delta f_3$
and observed a characteristic $k_\perp$ broadening away from the Gaussian form to a more exponential
decay. 
%
%
The relationship between  the genuine  and the  modified pCCFM  for unintegrated
parton distributions on the one hand and the evolution of integrated parton distributions on the other hand
is displayed in Fig.~\ref{relation}.
The modifications to the genuine  pCCFM equation  correspond to the Altarelli-Parisi part plus ladder contributions
in the evolution of the unintegrated parton distributions. 
The non-ladder contributions have to be 
added by hand. As the only difference in the evolution kernels for the integrated and the unintegrated
parton distributions in Fourier space is the factor $J_0(b_\perp q_\perp(1-z))$, which becomes unity as the transverse
impact parameter $b_\perp$ goes to zero, it is clear that the diagram between transverse integration and
evolution shown in Fig.~\ref{relation} commutes.
\newline
\newline
The next step will be to use the modified polarized CCFM equation  as
presented here to construct a consistent set of unintegrated polarized parton distributions
from existing data.
\section*{Acknowledgments}
This research was partially supported
by the EU Fourth Framework Program `Training and Mobility of Researchers',
Network `Quantum Chromodynamics and the Deep Structure of Elementary
Particles', contract FMRX--CT98--0194 and  by the Polish
Committee for Scientific Research (KBN) grants no. 2P03B 05119 and 5P03B 14420.
\section*{Appendix}
\subsection*{Approximation by Chebyshev polynomials}
A possible diagonalization of the problem from
a mathematical point of view would be the transformation of the pCCFM equation
(here using the weaker constraint $\Theta(Q-zq_{\perp})$)
into Mellin space:
\begin{eqnarray}
&& \overline{(\Delta P^{\rm CCFM}\otimes \Delta f)}( \omega ,b_\perp^2,Q^2) 
=
\int_0^1 dx x^{\omega-1} (\Delta P^{\rm CCFM}\otimes \Delta f)( x ,b_\perp^2,Q^2)
\nonumber \\
&=& \int_{Q_0^2}^{Q^2} \frac{d q_\perp^2}{q_\perp^2}\Delta \bar f(\omega, b_\perp^2, q_\perp^2)
\nonumber \\ &&\quad \times
\left( \int_0^1 dz z^{\omega-1} J_0((1-z){\bf |b_\perp|| q_\perp|})\Delta P^{\rm CCFM}(z,q_\perp^2, Q^2)
\right) 
\nonumber \\
\nonumber \\ && +
\int^{\infty}_{Q^2} \frac{d q_\perp^2}{q_\perp^2}\Delta \bar f(\omega, b_\perp^2, q_\perp^2)
\nonumber \\ &&\quad \times
\left(
\int_0^{\left(q_\perp^2/Q^2\right)^{-1/2}} dz z^{\omega-1} 
 J_0((1-z){\bf| b_\perp|| q_\perp|})\Delta P^{\rm CCFM}(z,q_\perp^2, Q^2)
\right) \;.
\nonumber \\
\end{eqnarray}
Unfortunately, it turns out that the back transformation after evolution, 
especially as regards the small-$x$ region, is numerically quite problematic.
Therefore, we take a different approach here, where both the $q_\perp$
and the $x$ dependence is expanded into properly chosen Chebyshev 
polynomials. These are defined in the following way:
\begin{eqnarray}
T_n(x) &=& \cos(n {\rm arccos}(x)) \nonumber \\ \\
T_0(x) &=&1 \nonumber \\
T_1(x) &=&x \nonumber \\
T_{n+1}(x) &=& 2x T_n(x) - T_{n-1}(x), \quad n\ge 1
\nonumber \\
\nonumber \\
\int_{-1}^{1} \frac{T_i(x) T_j(x)}{\sqrt{1-x^2}} dx &=& \left\{ \begin{array}{lll}
0 & \quad & i\ne j \\\\
\frac{\pi}{2} & \quad & i = j \ne 0 \\\\
\pi &\quad & i=j=0 \end{array} \right.\;.
\end{eqnarray}
The generic advantage of the Chebyshev polynomials is now that there exists also
a discrete orthogonality relation. Let
\begin{equation}
x^{(N)}_k := \cos\left[\pi \frac{\left(k-\frac{1}{2}\right)}{N}\right]
\end{equation}
be the $k$-th zero of $T_N(x)$ then one has as a discrete orthogonality relation
$(i,j<N)$:
\begin{equation}
\sum_{k=1}^{N} T_i(x_k) T_j(x_k) =  \left\{ \begin{array}{lll}
0 & \quad & i\ne j \\\\
\frac{N}{2} & \quad & i = j \ne 0 \\\\
N &\quad & i=j=0 \end{array} \right.\;.
\end{equation}
The central point is that for any arbitrary function f(x) in the interval
[-1,1] the $N$ coefficients $c_j$ given by:
\begin{eqnarray}
c_j &=& \frac{2}{N} \sum_{k=1}^N f(x^{(N)}_k) T_{j-1}(x^{(N)}_k) 
\nonumber \\
&=&  \frac{2}{N} \sum_{k=1}^N f \left[\cos\left(\pi 
\frac{k-\frac{1}{2}}{N}\right)\right]
\cos \left(\pi(j-1) 
\frac{k-\frac{1}{2}}{N}\right)\;.
\end{eqnarray}
yield an approximation formula:
\begin{equation}
f(x)\approx \left[\sum_{k=1}^{N}c_kT_{k-1}(x)\right]-\frac{c_1}{2}\;,
\end{equation}
which is exact on all $x^{(N)}_k, k=1,\dots,N$. Therefore, 
one can always substitute the continuous integral expression for the isolation
of the coefficients $c_j$ by a discrete one:
\begin{eqnarray}
c_j &=& \frac{2}{\pi}\int_{-1}^1 \frac{dx}{\sqrt{1-x^2}}T_{k-1}(x) f(x)
\nonumber \\
    &\to&\frac{2}{N} \sum_{k=1}^N f\left[ \cos\left(\pi 
\frac{k-\frac{1}{2}}{N}\right)\right]
\cos \left(\pi(j-1) 
\frac{k-\frac{1}{2}}{N}\right)\;.
\end{eqnarray}
So, choosing a cutoff $Q_{\rm max}$ large enough, one can define a set of variables:
\begin{eqnarray}
t' &=& 2\frac{\ln (q_\perp^2/Q_0^2)}{\ln (Q_{\rm max}^2/Q_0^2)}-1
\nonumber \\
t &=&  2\frac{\ln (Q^2/Q_0^2)}{\ln (Q_{\rm max}^2/Q_0^2)}-1\;.
\end{eqnarray}
Correspondingly for an $x_{\rm min}$ small enough one can define a second set of variables:
\begin{eqnarray}
y &=& 1-2\frac{\ln x}{\ln x_{\rm min}}
\nonumber \\
y' &=& 1-2\frac{\ln z}{\ln x_{\rm min}}
\nonumber \\
T_n(y') &=& T_n\left(1-2\frac{\ln z}{\ln x_{\rm min}}\right) = T^*_n(z)\;.
\end{eqnarray}
In this way one can expand:
\begin{equation}
\Delta \bar f(x,b_\perp, q_\perp^2) = \sum_{ij} \Delta \bar f_{ij}(b_\perp)c_iT_{i-1}(t')c_jT^*_{j-1}(x)\;,
\end{equation}
where $c_1 = 1/2$ and $c_k=1$ in all other cases.
In this way the CCFM equation transforms to a simple set of linear equations:
\begin{eqnarray}
\Delta \bar f( x ,b_\perp^2,Q^2) &=& \Delta
\bar f^0( x ,b_\perp^2)+(\Delta P^{\rm CCFM}\otimes \Delta \bar f)( x ,b_\perp^2,Q^2)
\nonumber \\
\Rightarrow \Delta \bar f_{ij}(b_\perp^2) &=& 
\Delta \bar f_{ij}^0(b_\perp^2) +  a_{ij\;i'j'}(b_\perp^2)\bar f_{i'j'}(b_\perp^2)
\nonumber \\
\nonumber \\
a_{ij\;i'j'}(b_\perp^2) &=& 
-\frac{\ln (x_{\rm min})\ln(Q^2_{\rm max}/Q_0^2)}{\pi^2}
c_{i'} c_{j'}\int_{-1}^{1} \frac{T_{i-1}(t) dt}{\sqrt{1-t^2}} 
                              \int_{-1}^1 \frac{T_{j-1}(y) dy}{\sqrt{1-y^2}}
\nonumber \\
&& \times  
\int_y^1 dy' 
\int_{-1}^{t-2(1-y')\frac{\ln x_{\rm max}}{\ln( Q^2_{\rm max}/Q^2_0)}} dt'
\nonumber \\ && \times
J_0((1-z){\bf| b_\perp|| q_\perp|})
\Delta P^{\rm CCFM}(z,t', t)
T_{i'-1}(t')
\nonumber \\ && \times
T^*_{j'-1}\left[\exp\left(\frac{\ln x_{\rm min}}{2}(y'-y)\right)\right]
\nonumber \\
\nonumber \\
 &\to& 
-\frac{\ln (x_{\rm min})\ln(Q^2_{\rm max}/Q_0^2)}{N^2}
c_{i'} c_{j'}
\Bigg\{\sum_{k=1}^N \sum_{k'=1}^N  T_{i-1}(t_k) T_{j-1}(y_{k'}) 
\nonumber \\
&& \times  
\int_{y_{k'}}^1 dy' 
\int_{-1}^{t_k-2(1-y')\frac{\ln x_{\rm max}}{\ln( Q^2_{\rm max}/Q^2_0)}} dt'
J_0((1-z){\bf| b_\perp|| q_\perp|})
\nonumber \\ && \times
\Delta P^{\rm CCFM}(z,q_\perp^2,t_k)
T_{i'-1}(t')T_{j'-1}(1-(y'-y_{k'}))\Bigg\}\;.  
\label{master}
\end{eqnarray}
Here we used: 
\begin{equation}
t_k=y_k = \cos\left(\pi\frac{k-\frac{1}{2}}{N}\right)\;.
\end{equation}
The generalization for the gluon and quark singlet matrix valued expression is straightforward.
The form of the master equation (\ref{master}) has now the advantage that one has to handle
only a simple two dimensional integration. For some contributions of the Altarelli-Parisi like type the
expressions are diagonal in $x$ and $q_\perp^2$. In those cases more simple expressions are possible:
\begin{eqnarray}
\Delta \bar f(x,Q^2,b_\perp^2) &=& \dots + \int_{Q_0^2}^{Q^2}\frac{dq_\perp^2}{q_\perp^2}
\frac{\alpha_s(q_\perp^2)}{2\pi} 
\nonumber\\ && \times \left[2+\frac{8}{3} \ln(1-x)\right]\Delta \bar  f(x,q_\perp,b_\perp^2)
\nonumber \\
\Rightarrow 
\sum_{kl} c_k c_l f_{kl}(b_\perp^2) T_{k-1}(t_x)T_{l-1}(t_{Q^2})
&=& \dots +\sum_{k'l'} c_{k'} c_{l'}  f_{k'l'}(b_\perp^2) 
\int_{Q_0^2}^{Q^2}\frac{dq_\perp^2}{q_\perp^2}
\frac{\alpha_s(q_\perp^2)}{2\pi} 
T_{l'-1}(t_{q_\perp^2})
\nonumber \\ && \times
 \left[2+\frac{8}{3} \ln(1-x)\right]T_{k'-1}(t_x)
\end{eqnarray}
\begin{eqnarray}
\Rightarrow  f_{kl}(b_\perp^2) &=& \dots + \sum_{l'k'}b_{kk'} d_{ll'} f_{k'l'}(b_\perp^2)
\nonumber \\
b_{kk'} &=& \sum_j c_{k'}\frac{2}{N} \left[2+\frac{8}{3}\ln(1-x_j)\right] T_{k'-1}(t_j) T_{k-1}(t_j) 
\;, \quad x_j = x_{\rm min}^{(1-t_j)/2}
\nonumber \\
d_{ll'} &=& \sum_j c_{l'}\frac{2}{N} T_{l-1}(t_j) 
 \int_{Q_0^2}^{Q^2_j}\frac{dq_\perp^2}{q_\perp^2}
\frac{\alpha_s(q_\perp^2)}{2\pi}  T_{l'-1}(t_{q_\perp^2})
\;, \quad Q^2_j = Q_{\rm max}^2 \left(\frac{Q^2_0}{  Q_{\rm max}^2}\right)^{(1-t_j)/2}\;,
\nonumber \\
\end{eqnarray}
where $t_j$ is again the $j$-th zero of $T_N(t)$.

\end{document}